\documentclass[12pt]{article}

\usepackage{epsfig,amsfonts,newlfont,rotate}

\DeclareMathAlphabet{\mathitbf}{T1}{cmr}{bx}{it}

\newcommand{\e}{\mathrm e}
\newcommand{\BPhi}{\mbox{\boldmath $\Phi$}}
\begin{document}

\title{Antiferromagnetic O($N$) models\\ in four dimensions}
\author{H.~G.~Ballesteros$^{a}$,
J.~M.~Carmona$^{b}$,\\
L.~A.~Fern\'andez$^{a}$ and
A.~Taranc\'on$^{b}$.}

\date{July 28, 1997}

\bigskip

\maketitle

\begin{center}
{\it a}~Departamento de F\'{\i}sica Te\'orica I, 
        Facultad de CC. F\'{\i}sicas,\\ 
{\it Universidad Complutense de Madrid, 28040 Madrid, SPAIN}\\
{\small e-mail: \tt hector, laf@lattice.fis.ucm.es}\\

{\it b}~Departamento de F\'{\i}sica Te\'orica,
Facultad de Ciencias,\\
{\it Universidad de Zaragoza, 50009 Zaragoza, SPAIN} \\
{\small e-mail: \tt carmona, tarancon@sol.unizar.es}\\
\end{center}

\bigskip

\begin{abstract}

We study the antiferromagnetic O($N$) model in the F$_4$ lattice.
Monte Carlo simulations are applied for investigating the behavior of the
transition for $N=2,3$.  The numerical results show a first order
nature but with a large correlation length.  The $N\to\infty$ limit is
also considered with analytical methods.

\end{abstract}
  
\vskip 10 mm

{\scriptsize

Monte Carlo.
Antiferromagnetic.
Phase transitions.
O($N$) models.


}
\vskip 5 mm

\newpage

\section{Introduction}

The antiferromagnetic formulations of field theories in four
dimensions have been recently paid considerable
attention~\cite{GALLAVOTTI, ISING, O4AF, RP2D4,POLONY}. The hope is to 
give some insight into the well known triviality problem in field
theory~\cite{TRIVIAL}. Also there are other interesting phenomena as
the apparition of new particles~\cite{POLONY}.

A spin model, in a simple cubic lattice with first neighbor
interactions, becomes antiferromagnetic if
the coupling is negative. However, with some exceptions \cite{RP2D4,RP2D3}, a
simple {\em staggered} transformation maps the antiferromagnetic phase
into the usual ferromagnetic one. 

To obtain a non-equivalent antiferromagnetic phase one has to include
further couplings or modify the lattice geometry (see for instance
ref \cite{O3AF}).

Perhaps the simplest method to obtain non-trivial antiferromagnetism
in four dimensions is to work in an F$_4$ lattice. It is defined by
taking out the odd sites (the sum of the coordinates is odd) of a simple
hypercubic lattice.

Four dimensional antiferromagnetic O($N$) models have been already 
studied by Monte Carlo(MC) means in this lattice. The O(1) model
(Ising model) was considered in reference \cite{ISING}:
a weak first order transition was found. A study of the O(4) model appears
in reference \cite{O4AF}: in the range of the lattice sizes simulated, 
the behavior pointed to a second order transition.

In this letter we consider the intermediate cases: O(2) and O(3), to
know if the order of the transition changes with $N$. We will
give evidence that the transitions are in both cases first order, but
the numerical difficulties grow with $N$. We also present an
analytical study in the $N\to\infty$ limit.

\section{The Model}

We label the coordinates of an F$_4$ lattice as a set of integers
$\{x,y,z,t\}$ such that $x+y+z+t$ is even. We consider the action

\begin{equation}
S = \beta H = -\beta\sum_{<i,j>} 
        \mathbf{\Phi}_i\cdot \mathbf{\Phi}_j\ ,
\label{ACCION}
\end{equation}
where the sum runs over 24 pairs of nearest neighbors and the field,
$\mathbf{\Phi}$, is a normalized $N$ component real vector.
We work in a hypercubic lattice of size $V=L^4/2$
with periodic boundary conditions. 
 
In the ferromagnetic region ($\beta>0$) this model is expected to
belong to the same universality class of the simple cubic model: it
presents a second order transition with mean field critical exponents.

In the antiferromagnetic sector ($\beta<0$) the system also presents
an ordering phase transition, but the structure of the ordered phase
is much more complex. In fact, it can be easily checked that the 
ground state presents frustration. 
Moreover, the ordered vacuum is not isotropic.

The independent order parameters we can construct with periodicity 2 are

\begin{equation}
\begin{array}{lcl}
\mbox{\boldmath $M$}_\mathrm{F}&=&\displaystyle 
       \frac{1}{V} \sum_{x,y,z,t}\mathbf{\Phi}_{xyzt} ,\\
\mbox{\boldmath $M$}_\mathrm{AFH}^x&=&\displaystyle 
       \frac{1}{V} \sum_{x,y,z,t}\mathbf{\Phi}_{xyzt}(-1)^x ,\\
&\vdots\\
\mbox{\boldmath $M$}_\mathrm{AFP}^{x-y}&=&\displaystyle 
       \frac{1}{V} \sum_{x,y,z,t}\mathbf{\Phi}_{xyzt}(-1)^{x+y} ,\\
&\vdots\\
\end{array}
\label{PARORD}
\end{equation}
where the sums are extended to all the F$_4$ sites.
The dots stand for the other 3 combinations of hyperplanes and 2
of planes.

We label the site $I=1,\ldots,8$ inside the $2^4$ elementary cell
with its Cartesian coordinates $X,Y,Z,T=0,1$.
In practice we measure the 8 different magnetizations associated to a
given position in the elementary cells, $\mbox{\boldmath $m$}_i$,
defined as the normalized sum of the magnetization for all $2^4$ cells
for each of the 8 sites.

\begin{equation}
\mbox{\boldmath $m$}_I=\mbox{\boldmath $m$}_{XYZT}=\displaystyle 
       \frac{8}{V} \sum_{x,y,z,t \atop 
x-X,\ldots\ \mathrm{even}}
        \mbox{\boldmath $\Phi$}_{xyzt}\ .
\end{equation}

The quantities (\ref{PARORD}) can be expressed as linear combinations 
of these magnetizations.

The mean magnetization and the susceptibility are defined respectively as:
\begin{equation}
M=\left< \sqrt{\frac{1}{8}\sum^{8}_{I=1} \mbox{\boldmath $m$}^2_I} \right>,
\quad \chi=\frac{V}{8}
\left< \sum^{8}_{I=1} \mbox{\boldmath $m$}^2_I \right>.
\end{equation}

The Binder Cumulants are defined in a such way 
that $\left.V_M\right|_{\beta=0}=0$ and $\left.V_M\right|_{\beta=\infty}=1$:

\begin{equation}
V_M^{\mathrm{O}(N)}=\frac{N+2}{2} \left( 1-\frac{
        8N\left< \sum^8_{I=1} (\mbox{\boldmath $m$}^2_I)^2 \right> }
        {(N+2) {\left< \sum^8_{I=1}   \mbox{\boldmath $m$}^2_I 
        \right> }^2 }
        \right).
\end{equation}

For the connected susceptibility we use the definition
\begin{equation}
\chi_{\mathrm{c}}=V\left(\frac{1}{8} \sum^8_{I=1} 
        \left< \mbox{\boldmath $m$}^2_I \right> - M^2 
        \right).
\end{equation}

\section{Critical behavior}

For an operator $O$ that diverges as $(\beta-\beta_\mathrm{c})^{-x_O}$,
its mean value at a coupling $\beta$ in a size $L$
lattice can be written, in the critical region, assuming the finite-size
scaling ansatz as~\cite{LIBROFSS} 
\begin{equation}
O(L,\beta)=L^{x_O/\nu}\left(F_O(\xi(L,\beta)/L)+O(L^{-\omega})\right), 
\label{FSS}
\end{equation}
where $F_O$ is a smooth scaling function and $\omega$ is the
universal leading corrections-to-scaling exponent. 
In order to eliminate the unknown $F_O$ function 
we can measure at the coupling where 
$F_O$ presents a maximum as for $\chi_{\mathrm c}$ or for the
specific heat $C_V$.
Another method~\cite{RP2D3} is to 
study the behavior of the quantities

\begin{equation}
Q_O=O(sL,\beta)/O(L,\beta).
\end{equation}
We use $V_M L$ as scaling variable~\cite{RP2D3}. It is direct to obtain 
\begin{equation}
\left.Q_O\right|_{Q_{V_M L}=s}=s^{x_O/\nu}+O(L^{-\omega}).
\label{QUO}
\end{equation}
We use that $x_{\chi}=\gamma$ and  
$x_{ \partial_{\beta} \log (M)}=1$ to obtain the critical exponents.

The expected FSS behavior of a first-order transition~\cite{FSSFIRST}
corresponds to apparent exponents: $\nu=1/d,\ \alpha=1,\ \gamma=1$.

\section{The Simulation}

We will consider in this letter the cases $N=2,3$. We have used a Metropolis
algorithm followed by $N_\mathrm{o}$ overrelaxation steps as update 
method, $N_\mathrm{o}$ depending on the model and lattice size.
We have worked in lattice sizes up to 48. 
We used for the computation the dedicated machine RTNN, consisting in
16 Dual Pentium Pro units. For the largest lattices we parallelized,
using shared memory, in each dual motherboard.
Every $f$ sweeps we measured the energy, the specific heat
and the 8 period-two magnetizations $\mbox{\boldmath $m$}_I$. 
In table~\ref{SIMULATIONS} we report the parameters of the simulation
at the critical region.

In order to extrapolate to the neighborhood of the critical point, we
used the usual reweighting method~\cite{FERRSWEN}.
We also simulated at the maxima of the specific heat 
because the region where the extrapolation was reliable 
was not large enough. These simulations were about one fourth of the
total CPU time.

The errors were computed with a jack-knife method, performing 50 blocks
in order to achieve statistical error bars within 10\% of precision. 

\begin{table}[t]
\begin{center}
\begin{tabular}{|r|c|c|c|r|l|}\hline
\multicolumn{1}{|c|}{$L$} 
    & \multicolumn{1}{c|}{$f$} 
    & \multicolumn{1}{c|}{$N_\mathrm{o}$}      
    & \multicolumn{1}{c|}{$\tau(\chi)$}      
    & \multicolumn{1}{c|}{\# of $\tau$}      
    & \multicolumn{1}{c|}{$\beta$}\\ \hline\hline
4 &      20&      3&          0.775(9) & 100000 & -0.352\\\hline
6 &      20&      3&          1.144(14)& 87000  & -0.352\\\hline
8 &      20&      3&          1.63(4)  & 55000  & -0.352\\\hline
12&      20&      3&          3.16(6)  & 60000  & -0.352\\\hline
16&      20&      3&          6.5(5)   & 11400  & -0.352\\\hline
24&      25&      4&          15.8(7)  & 5000   & -0.3516\\\hline
32&      24&      7&          32(1)    & 1700   & -0.3513\\\hline 
48&      32&      7&          85(7)    & 312    & -0.35125\\\hline\hline\hline
4 &      20&      3 &         0.871(11) &     98000   &       -0.53  \\\hline
6 &      20&      3 &         1.19(2)   &     82000   &      -0.53   \\\hline
8 &      20&      3 &         1.52(3)   &     98700   &      -0.53   \\\hline
12&      20&      3 &         2.07(4)   &     63000   &      -0.53   \\\hline
16&      20&      3 &         3.15(6)   &     48000   &      -0.53   \\\hline
24&      24&      5 &         4.99(14)  &     29000   &      -0.5287 \\\hline
32&      28&      6 &         8.7(4)    &     5000    &      -0.5287 \\\hline
48&      28&      6 &         21(1)     &     2000    &      -0.5286 \\\hline

\end{tabular}
\caption{Description of the simulation in the critical region for O(2)
(upper part) and O(3) (lower part). We report the lattice size, the
frequency of measures, number of over-relaxation steps for each
Metropolis one, autocorrelation time ($\tau$) for $\chi$, iterations in $\tau$
units, and the coupling.}
\label{SIMULATIONS}
\end{center}
\end{table}

\section{The Vacuum}

For large $|\beta|$, all the magnetizations $\mbox{\boldmath $m$}_I$
go to unitary vectors, confirming the assumption of a period 2 vacuum,
so that we can restrict our analysis only to a unit cell.  
We also find that $\mbox{\boldmath $M$}_\mathrm{AFP}$ is non zero,
while $\mbox{\boldmath $M$}_\mathrm{F}$ and $\mbox{\boldmath
$M$}_\mathrm{AFH}$ vanish.

When $\mbox{\boldmath $M$}_\mathrm{F}=\mbox{\boldmath $M$}_\mathrm{AFH}=0$ ,
 it follows that
$\mbox{\boldmath $m$}_{XYZT}=\mbox{\boldmath
$m$}_{1-X,1-Y,1-Z,1-T}$.
We checked that the cosine between 
$\mbox{\boldmath $m$}_{XYZT}$ and 
$\mbox{\boldmath$m$}_{1-X,1-Y,1-Z,1-T}$ goes to 1, with 
the expected $L^{-4}$ behavior in the broken phase corresponds to AFP order.

This ordering is also found near the transition. 
So we can restrict ourselves to study only 4 independent spins.

At $T=0$ the frustrated ground state is described in \cite{O4AF}; it
consists in two couples of antiparallel spins, but the angle between
couples is free.  The different choices for the couples determine the
plane for AFP symmetry breaking.

At $T>0$ we do not know whether there is a privileged angle or not. 
It is necessary to determine the pattern of the
symmetry breaking.

A very important point is to determine if the two couples 
are aligned or not.
To clarify this point we construct the following tensor:
\begin{equation}
T^{ab}=\frac{1}{4}\sum^4_{K=1} 
         \mbox{\boldmath $m$}_K^a \times
                 \mbox{\boldmath $m$}_K^b ,    
\label{TENSOR}
\end{equation}
where the superindices run for the components of the $N$ vectors and 
the sum over the four independent magnetizations in the elementary cell.
It is clear that if the two couples are aligned, the four tensors
as well as the sum  
tensor can be simultaneously diagonalized, and so we expect
in the broken phase a non-zero value for the largest eigenvalue 
and zero values (up to $L^{-4}$ effects) for the rest ($N-1$) of them. 

If this holds true, we also expect in the critical region 
a $L^{-2\beta/\nu}$ behavior  for the biggest one and 
a $L^{-4}$ for the others. This will be checked in the next section.

\begin{table}[t]
\begin{center}
\begin{tabular}{|c|c|c|}\hline
\multicolumn{1}{|c|}{Eigenvalue}      
&\multicolumn{1}{c|}{$\chi^2/{\mathrm {d.o.f.}}$}
&\multicolumn{1}{c|}{Value}\\ \hline\hline
$\lambda_\mathrm{max}$&0.16&0.20346(15)\\\hline
$\lambda_\mathrm{min}$&0.13&-0.0(2.2)$\times 10^{-6}$\\\hline\hline
$\lambda_\mathrm{max}$&0.11&0.18986(7)\\\hline
$\lambda_\mathrm{med}$&0.89&2.3(2.7)$\times 10^{-7}$\\\hline
$\lambda_\mathrm{min}$&0.05&-1.7(1.2)$\times 10^{-6}$\\\hline
\end{tabular}
\end{center}
\caption{Eigenvalues for the tensor (\ref{TENSOR}) for the O(2) 
model (upper part) at $\beta=-0.37$ and for the O(3) model (lower part) 
at $\beta=-0.57$.}
\label{EIGEN}
\end{table}

In table~\ref{EIGEN} we show the eigenvalues for both  models
in the broken phase. We note that the largest one goes to a non-zero 
value while the rest go to zero. The fit parameters are obtained with
a linear extrapolation in $L^{-2}$ for the former case 
and in $L^{-4}$ for the latter. 

\section{Critical exponents}

A determination of the critical exponents is obtained
by studying
the height of the peaks of  $C_V$ and $\chi_\mathrm{c}$.
For a first order transition both quantities should diverge
as the volume ($\alpha/\nu=4, \gamma/\nu=4$). For small lattices
it is usual to find the apparent critical exponents of a {\em weak
first order transition}: $\alpha/\nu=1, \gamma/\nu=1$ (see
ref.~\cite{WEAK}), which are precursors of a first order transition.

\begin{figure}[t]
\begin{center}
{\centering\epsfig{file=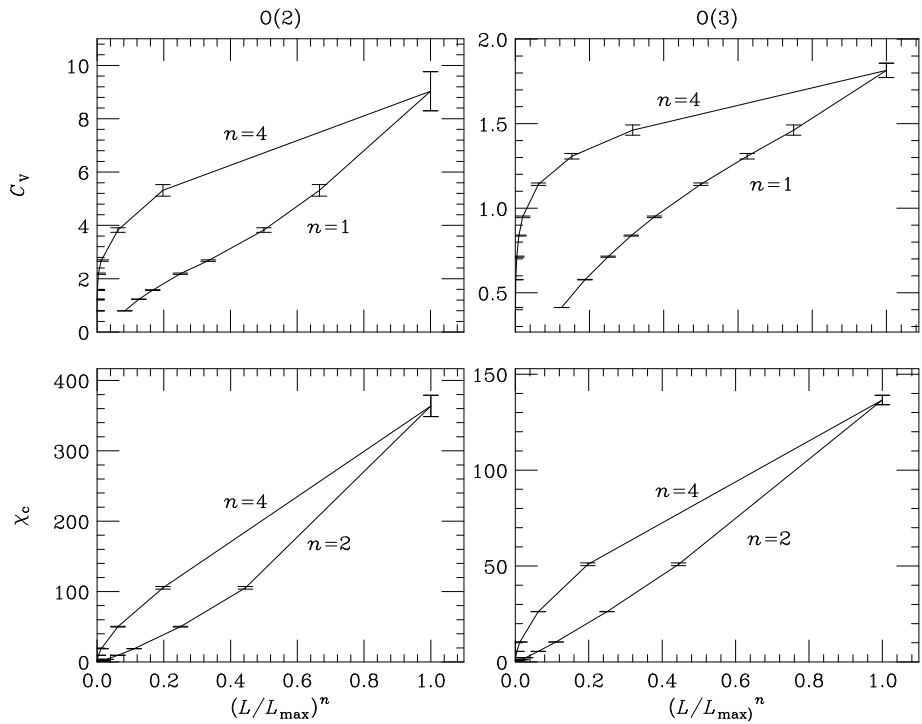,width=0.7\linewidth,angle=90}}
\end{center}
\protect\caption{Specific heat and connected susceptibility for O(2)
and O(3) as a function of several powers of the lattice sizes.}
\label{POWERS}
\end{figure}

To analyze the divergence of $C_V$ a bilogarithmic plot is
not adequate due to the presence of a non-negligible constant term.
In order to compare with the first order behavior it is better to plot
$C_V$ and $\chi_\mathrm{c}$ as a function of several powers
of $L$. This is done in fig. 1.
We remark that in the O(2) case for $L$ in the interval [8,24] ( [12,32]
for O(3)) there is an excellent linear fit for $n=1$ which is the
value predicted in a weak first order transition. This frequently produces a
misunderstanding of the order of the transition. However, it is clear
from fig.~\ref{POWERS} that this is a transient effect: for the larger
lattices the divergences are faster than linear, and presumably they
would reach the first order behavior for very large lattices.

The susceptibility also shows a fast divergence. Although we are not
able to observe the asymptotic first order behavior the trend seems
rather clear.

\begin{figure}[t]
\begin{center}
{\centering\epsfig{file=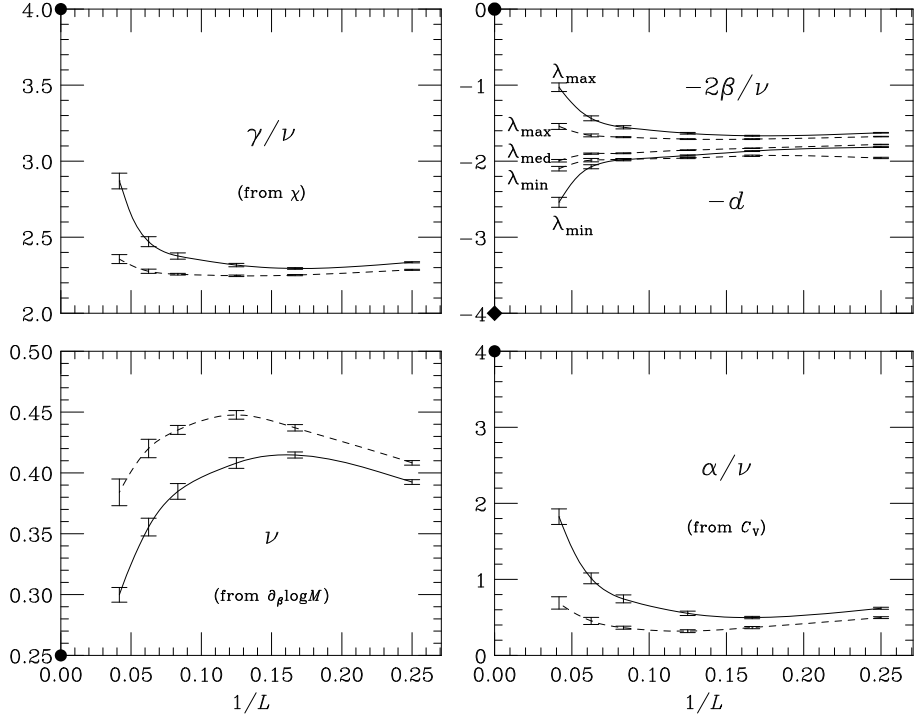,width=0.7\linewidth,angle=90}}
\end{center}
\protect\caption{Critical exponents for O(2) (solid lines) and O(3) 
(dashed ones) measured through the relation (\ref{QUO}). The filled
circles (diamond for the non-maximum eigenvalues) mark the first order
limit.}
\label{EXPO}
\end{figure}

A more accurate measure of the critical exponents can be obtained
from eq.~(\ref{QUO}). We have always used the ratio $s=2$.
In fig.~\ref{EXPO} we plot several determinations of exponents 
using different operators. 
We remark that there is a systematic error in the $\alpha/\nu$
determination because of the analytic term in $C_V$.
We observe no asymptotic behavior in
all cases although the values in the larger lattices are hardly
compatible with a second order transition.

In the upper right part of fig.~\ref{EXPO} we plot the exponents
related with each of the eigenvalues of the matrix (\ref{TENSOR}).
While the maximum eigenvalues should behave as $L^{-2\beta/\nu}$ (with
$\beta/\nu=0$ for a first order transition), the others should go
to zero as $L^{-d}$. The latter can be used as a control of
when the asymptotic regime is reached. We observe that we are far
from this regime but the first order limit seems rather clear.

A comparison between the curves for O(2) and O(3) shows a roughly
similar shape, differing in a horizontal shift that corresponds to
multiplying the lattice size by a factor near 2. This fact can be
understood as a correlation length at the critical point which is
twice larger for O(3) than for O(2).

\section{Critical point and energy histograms}

To calculate an estimation for the critical coupling 
we study the  $\beta_L$  values where 
$V_M(2L,\beta_L)=V_M(L,\beta_L)$. In both models, 
these $\beta_L$ for the largest lattices are compatible within
the error bars. 
We can fit to the functional form $\beta_{\mathrm c}^L-
\beta_{\mathrm c}(\infty)\propto L^{-x}$
in order to estimate the error bars.
We perform these fits with the full covariance matrix. In both cases we
obtain fitting for $L\geq 6$ a wide valid range for $x$ and very good 
$\chi^2$ ($\chi^2/\textrm{d.o.f.}=1.9/2$ for  O(2) and  
$\chi^2/\textrm{d.o.f.}=0.8/2$ for O(3)).
The results are  compatible with the values for the largest lattices.
We get 
\begin{eqnarray}
{\beta_{\mathrm c}(\infty)}^\mathrm{O(2)}&=&-0.351216(10)\\
{\beta_{\mathrm c}(\infty)}^\mathrm{O(3)}&=&-0.52857(2).
\end{eqnarray}

\begin{figure}[t]
\begin{center}
{\centering\epsfig{file=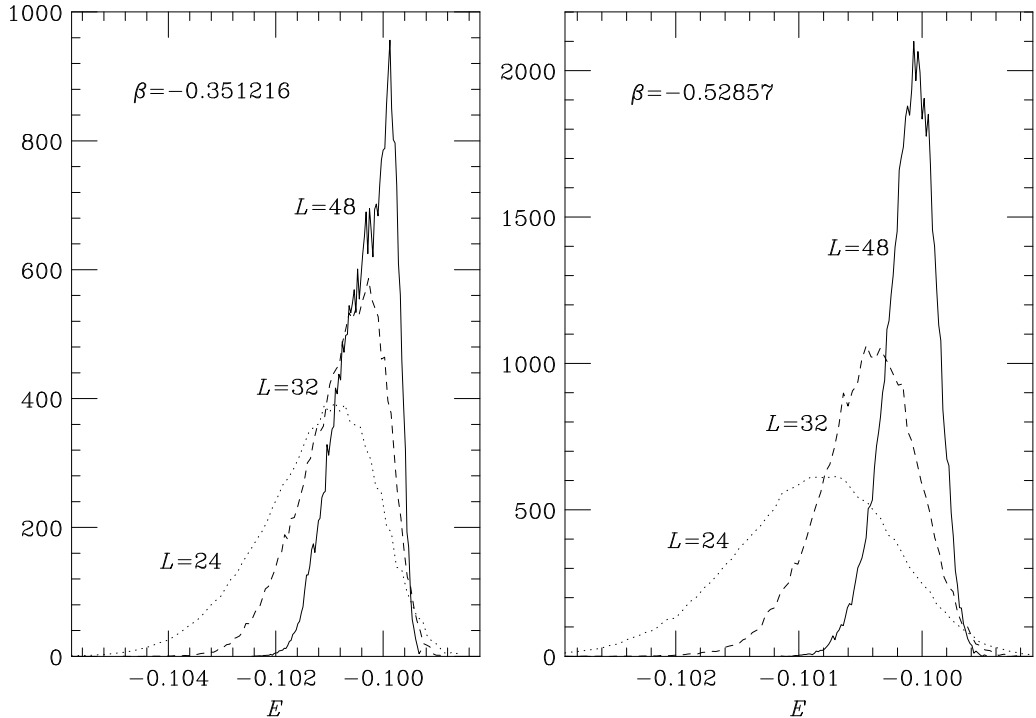,width=0.7\linewidth,angle=90}}
\end{center}
\protect\caption{Normalized energy histogram at $\beta_{\mathrm c}(\infty)$ 
for both models. }
\label{HISTO}
\end{figure}

Finally let us comment on the energy distribution of the
configurations. A direct check of the first order character of a
transition is the observation of a latent heat. Unfortunately,
a sharp double peak structure can be observed only when the lattice
size is much larger than the correlation length at the critical point.
In figure~\ref{HISTO} we show the energy histograms for 
both models at $\beta_{\mathrm c}$.  
In the O(2) case we note that the width of the energy distribution is
nearly constant for the larger lattices, being an indication of the 
existence of a two peak distribution that cannot be resolved. 
In the O(3) case, up to $L=48$ there is not a similar behavior.

\section{The {\boldmath $N\to\infty$} limit}

The partition function of the model can be written as
\begin{eqnarray}
Z &=& Z_0 \int \prod_j d^N\BPhi_j \, \delta(\BPhi_j^2-1)\ \e^{-N\beta' H}
\nonumber \\
&=&Z_0 \int \prod_j d^N\BPhi_j {N \, d\alpha_j\over 2\pi} \e^{N \left\{
\sum_j ({\mathrm i}\alpha_j+\lambda_j)(1-\BPhi_j^2)-\beta' H
\right\}},
\label{PARTITION}
\end{eqnarray}
where $\beta'=\beta/N$, and $Z_0$ is a normalization factor 
such that $Z\to 1$ when 
$\beta\to 0$. 
We have introduced the conjugate parameters $\alpha_j, \lambda_j$ to
give an integral representation of the constraint
$\BPhi_j^2=1$~\cite{ITZYKSON}.

Writing the quadratic form in the exponent of eq. (\ref{PARTITION}) as
\begin{equation}
{-1\over 2} \sum_{n,m} \phi_n Q_{nm} \phi_m ,
\end{equation}
the integration over $\BPhi$ yields
\begin{equation}
Z=Z_0 \left( {2\pi\over N}\right)^{1/2(N-2)V}\int \prod_j d\alpha_j \,
\e^{N/2 \left\{ \sum_j 2(\lambda_j+{\mathrm i}\alpha_j)- \mathrm{Tr} 
\ln Q\right\}}.
\end{equation}

In the limit $N\to\infty$, a variational equation with respect to
$2(\lambda_k+i\alpha_k)$ gives
\begin{equation}
1=(Q^{-1})_{ii}.
\label{VARIA}
\end{equation}

In order to study the disorder-AF transition in the F$_4$ lattice ($\beta<0$),
we perform a change of variables which transforms the plane-AF vacuum 
(suppose $x-y$) into a ferromagnetic one defining 
\begin{equation}
\BPhi'_{xyzt}=(-1)^{x+y}\BPhi_{xyzt}.
\end{equation}

$Q$ matrix changes and then the propagator, $Q^{-1}$, reads

\begin{equation}
G(p)={1\over(-\beta')}
{1 \over \xi^{-2}+4(2+g(p))}, 
\end{equation}
where
\begin{eqnarray}
g(p)&=&\cos p_x\cos p_y+\cos p_z\cos p_t-
\cos p_x\cos p_z \\\nonumber
&&-\cos p_x\cos p_t-\cos p_y\cos p_z-\cos p_y\cos p_t,
\end{eqnarray}
and  $\xi$ is defined from (translationally invariant)
auxiliary fields as
\begin{equation}
2(\lambda_i+{\mathrm i}\alpha_i)\equiv (-\beta')(\xi^{-2}+8).
\end{equation}
From the variational equation (\ref{VARIA}) we obtain $\beta'_{\mathrm c}$
imposing $\xi=\infty$:
\begin{equation}
\beta_c'=\int {d^4p \over (2\pi)^4} {1\over -8-4g(p)}=-0.178972.
\label{BETACN}
\end{equation}

\begin{table}[t]
\begin{center}
\begin{tabular}{|l|c|}\hline
\multicolumn{1}{|c|}{$N$}      
&\multicolumn{1}{c|}{$\beta_\mathrm{c}$}\\ \hline\hline
1&-0.17459(15)\\\hline
2&-0.175608(5)\\\hline
3&-0.17619(1)\\\hline
4&-0.1766(1)\\\hline
\end{tabular}
\caption{Critical couplings divided by $N$ for different O($N$) models.
We include also the obtained in references \cite{ISING} and \cite{O4AF}.}
\label{BETACR} 
\end{center}
\end{table}

In table~\ref{BETACR} we compare (\ref{BETACN}) with MC results for
$N=1,2,3,4$. The good agreement between the simulations for these values
of $N$ and the analytical limit when $N\to\infty$ points to the absence of an
abrupt change of the critical properties as a function of $N$. Exactly
at $N=\infty$ the order of the transition is not clear, because the
divergence in the correlation length can simply be caused by the Goldstone
bosons of the symmetry breaking.

\section{Conclusions}

In this letter we present a MC study of the four dimensional
antiferromagnetic O(2) and O(3) models in the F$_4$ lattice. We study
the critical behavior of these models with FSS
techniques. There is an apparent asymptotic behavior which gives false
critical exponents for not large enough lattice sizes. This transitory
effect can be understood as caused by a large correlation length whose
presence can be demonstrated for some observables (as the eigenvalues
of the sum tensor of the period-two magnetizations). 
This must be very carefully controlled, because as we see in our
case, the behavior changes drastically when larger lattice sizes are
considered, revealing the true first order nature of the O($2$), O($3$)
transitions. We also see that this effect is bigger as $N$ grows, so
that for larger values of $N$, it is very difficult to study 
numerically the critical
properties of the system. However, the great accuracy in the determination
of the critical point obtained by the analytical calculation at $N=\infty$  
points to a similar qualitative behavior for all values of $N$.

\section*{Acknowledgments}

We thank to the CICyT (contracts AEN94-0218, AEN96-1634) for partial
financial support. We have employed for the simulations dedicated Pentium
Pro machines (RTNN project). J.M. Carmona is a Spanish MEC fellow. 

\hfill

\end{document}